\documentstyle[twocolumn,prl,aps,epsf]{revtex}

\begin{document}


\def\isn{$^4$}
\def\sphn{$^1$}
\def\md{$^3$}
\def\jlab{$^5$}
\def\mit{$^7$}
\def\rutgers{$^6$}
\def\bale{$^2$}
\def\lns{$^{11}$}
\def\ncat{$^{10}$}
\def\fiu{$^{8}$}
\def\ipn{$^{12}$}
\def\yerevan{$^{9}$}

\title{A precise measurement of the deuteron elastic structure function 
	A(Q$^{\hbox{\bf 2}}$)}

\author{D.~Abbott,\jlab\   
	A.~Ahmidouch,\mit$^,$\ncat\  
	H.~Anklin,\fiu\   
	J.~Arvieux,\lns$^,$\ipn\   
	J.~Ball,\sphn$^,$\lns\   
	S.~Beedoe,\ncat\   
	E.J.~Beise,\md\   
	L.~Bimbot,\ipn\    
	W.~Boeglin,\fiu\   
	H.~Breuer,\md\   
	R.~Carlini,\jlab\    
	N.S.~Chant,\md\   
	S.~Danagoulian,\ncat\   
	K.~Dow,\mit\   
	J.-E.~Ducret,\sphn\    
	J.~Dunne,\jlab\
	R.~Ent,\jlab\    
	L.~Ewell,\md\   
	L.~Eyraud,\isn\   
	C.~Furget,\isn\   
	M.~Gar\c con,\sphn\    
	R.~Gilman,\jlab$^,$\rutgers\   
	C.~Glashausser,\rutgers\   
	P.~Gueye,\jlab\   
	K.~Gustafsson,\md\   
	K.~Hafidi,\sphn\    
	A.~Honegger,\bale\    
	J.~Jourdan,\bale\   
	S.~Kox,\isn\   
	G.~Kumbartzki,\rutgers\   
	L.~Lu,\isn\    
	A.~Lung,\md\   
	D.~Mack,\jlab\   
	P.~Markowitz,\fiu\   
	J.~McIntyre,\rutgers\   
	D.~Meekins,\jlab\   
	F.~Merchez,\isn\    
	J.~Mitchell,\jlab\
	R.~Mohring,\md\    
	S.~Mtingwa,\ncat\   
	H.~Mrktchyan,\yerevan\   
	D.~Pitz,\sphn$^,$\md$^,$\jlab\    
	L.~Qin,\jlab\    
	R.~Ransome,\rutgers\   
	J.-S.~R\'eal,\isn\   
	P.G.~Roos,\md\    
	P.~Rutt,\rutgers\    
	R.~Sawafta,\ncat\   
	S.~Stepanyan,\yerevan\      
	R.~Tieulent,\isn\   
	E.~Tomasi-Gustafsson,\sphn$^,$\lns\    
	W.~Turchinetz,\mit\   
	K.~Vansyoc,\jlab\    
	J.~Volmer,\jlab\   
	E.~Voutier,\isn\   
	W.~Vulcan,\jlab\    
	C.~Williamson,\mit\   	
	S.A.~Wood,\jlab\    
	C.~Yan,\jlab\   
	J.~Zhao,\bale\   
	and 
	W.~Zhao\mit\\
        (The Jefferson Lab t$_{\hbox{20}}$ collaboration)
       }

\address{
\sphn DAPNIA/SPhN, CEA/Saclay, 91191 Gif-sur-Yvette, France\\ 
\bale Basel Institut fur Physik, Switzerland\\
\md University of Maryland, College Park, MD 20742, USA\\
\isn ISN, IN2P3-UJF, 38026 Grenoble, France\\
\jlab Thomas Jefferson National Accelerator Facility, Newport News, VA 23606, USA\\
\rutgers Rutgers University, Piscataway, NJ 08855, USA\\
\mit M.I.T.-Bates Linear Accelerator, Middleton, MA 01949, USA\\
\fiu Florida International University, Miami, FL 33199, USA\\
\yerevan Yerevan Physics Institute, 375036 Yerevan, Armenia\\
\ncat North Carolina A. \& T. State University, Greensboro, NC 27411, USA\\
\lns LNS-Saclay,  91191 Gif-sur-Yvette, France\\
\ipn IPNO, IN2P3, BP 1, 91406 Orsay, France\\
}

\date{October 28, 1998}

\maketitle

\begin{abstract}
The $A(Q^2)$ structure function in elastic electron-deuteron  scattering 
was measured at six momentum transfers $Q^2$ between 0.66 
and 1.80 (GeV/c)$^2$ in Hall C at Jefferson Laboratory.
The scattered electrons and recoil deuterons were detected in coincidence,
at a fixed deuteron angle of 60.5$^{\circ}$. 
These new precise measurements resolve discrepancies
between older sets of data. They put significant constraints
on existing models of the deuteron electromagnetic structure,
and on the strength of isoscalar meson exchange currents. 
\end{abstract}

\narrowtext

The deuteron is the only two nucleon bound state, and as such
 is one of the most fundamental systems in nuclear physics. 
Measurements of its electromagnetic properties 
have been invaluable to our understanding of the nucleon--nucleon
interaction and of the role of meson and isobar degrees of freedom in nuclear
systems. At intermediate to high momentum transfer, it remains a challenge
to explore the limitations of the meso-nucleonic picture of nuclei, and 
unravel the possible role of the quark substructure of nucleons in 
nuclear structure. The deuteron electromagnetic form factors as measured 
in elastic $ed$ scattering provide a crucial test for any  
model of the deuteron.
In this paper, new measurements of the deuteron elastic structure function 
$A(Q^2)$ are presented, 
in an intermediate momentum transfer region where previous 
experiments \cite{ARN75,CRA85,ELI69,PLA90}
differ by as much as 40\% from each other and where theoretical models
 have been recently refined.
   
Assuming single photon exchange,
the electron-deuteron unpolarized elastic differential cross-section
can be written as :
\begin{equation}
\frac{d\sigma}{d\Omega} = 
\sigma_{NS}  
\left[ A(Q^2) + B(Q^2) \tan^2\frac{\theta_e}{2} \right]
\end{equation}
where 
$\sigma_{NS}$ 
is the  Mott cross-section
multiplied by the deuteron recoil factor \cite{Mott},  
$Q$ the four-momentum transfer and 
$\theta_e$ the  electron scattering angle.
$A(Q^2)$ and $B(Q^2)$ are two structure functions, quadratic
combinations of the three electromagnetic form factors
(charge monopole, quadrupole and magnetic dipole)
which characterize a spin 1 nucleus : 
\begin{equation}
A(Q^2) = G^2_C(Q^2) + \frac{8}{9} \tau^2 G^2_Q(Q^2) 
		    + \frac{2}{3} \tau G^2_M(Q^2),
\end{equation}
\begin{equation}
B(Q^2) = \frac{4}{3} \tau (1+\tau) G^2_M(Q^2),
\end{equation}
with $\tau = Q^2/4M_d^2$.
In the kinematic conditions of the present experiment, 
$A$ is dominated by the contribution from $G_Q$,
and to a lesser extent by the one from $G_C$, which exhibits 
a node \cite{GAR94,FUR98}. 
Existing measurements of $B$ \cite{BOS90} indicate that 
its contribution to our forward angle cross-sections
is always smaller than 1.6$\%$.

 The two photon exchange contribution,
where a virtual photon couples to each nucleon,
has been estimated to contribute
up to a few \% to the $ed$ cross-sections \cite{2PH}. Because of 
uncertainties coming from approximations in the estimate, this 
double scattering contribution
is neglected here, but deserves more investigation.

The measurements were performed in Hall C at
Jefferson Laboratory, as part of an experiment devoted to
the determination of the deuteron tensor polarization \cite{FUR98},
with specific conditions for the precise
determination of absolute cross-sections.
An 80 $\mu$A continuous electron beam was used 
 on a 4.45 cm thin-walled aluminium cell filled with liquid deuterium.
The target cryogen  contained 99\% deuterium and
1\% hydrogen.
The beam spot was rastered on the target ($\pm$ 1 mm in both
directions), and the density reduction due to beam heating
was measured to be about 1.3\%.
The integrated beam charge was measured 
with three resonant cavity monitors and a parametric current transformer.
The beam energies, between 1412 and 4050 MeV, were measured  
to 0.1\%.
The scattered electrons were detected in the 
Hall C High Momentum Spectrometer (HMS) \cite{HMS}. 
This QQQD spectrometer is equipped 
with a detector package of two drift chambers (6 planes each), 
two scintillator hodoscopes (2 planes each), an electromagnetic calorimeter
and a gas \v{C}erenkov detector, resulting in an electron detection
efficiency of 94-99\%.

For a precise definition of the electron solid angle
and
to minimize the acceptance mismatch with the recoil deuteron detector, 
the HMS was equipped with a specially designed tungsten collimator
($\pm$ 8.01 mrad in horizontal, $\pm$ 43.8 mrad in vertical).
The resulting solid angle, as determined from a Monte-Carlo simulation,
was  1.386 msr. 
The recoil deuterons were detected in coincidence with the
electrons using 
the upstream two  scintillators of the POLDER polarimeter \cite{KOX94},
after passing through a fixed angle (60.5$^{\circ}$) magnetic channel.
The deuteron channel (DC) was a QQSQD design 
with a point to point focus in the
vertical plane, and zero magnification and kinematics dependent focussing
in the horizontal dispersive plane. 
Multiple scattering and absorption in the DC were small compared to
that in the target fluid. The small acceptance mismatch between
the HMS and the DC (1 to 3.7\%) was modeled at the lowest three
kinematic points, and measured at the highest three points.
The elastic $ed$ events were identified unambiguously through the
HMS determination of the electron
momentum and angle, the $ed$ coincidence timing peak, and the deuteron energy
loss in the
POLDER scintillators.
Target wall contributions (which were subtracted) were approximately 0.1\%.

In the data analysis, a cut at -4\% was applied 
on the electron momentum relative to the elastic peak.
The corresponding corrections (about 20\%) due to
losses in the radiative tail were calculated 
according to Ref. \cite{MT69}. 
The deuteron losses through nuclear scattering and absorption  
were estimated from measured and
calculated deuteron-nucleus ($dA$) total cross-sections. Calculations, 
necessary in the case of $dd$ scattering where no data are available, 
were based on the Glauber formalism following Ref.\cite{CHA90}.  
These losses amounted to 3.2 to 6.2\%, depending
on the kinematical setting.

Our results are shown in Table 1. 
The systematic errors on $A(Q^2)$ come from uncertainties in 
the beam current measurements (0.5\%), 
the beam energy (about 1\%),
the target length and density (1.3\%), 
the electron angle (2.3-4.7\%) 
and solid angle (1\%), 
the electron tracking efficiency (0.2-1.2\%),
the radiative corrections (1.5\%), 
the mismatch between DC and HMS (0.5-1\%), 
and the deuteron losses (0.6-1.2\%). These were combined quadratically.
The small $B(Q^2)$ contribution was subtracted using a fit to the world data,
with no additional contribution to the systematic errors.
Elastic electron-proton ($ep$)
cross-sections 
were measured for all six energies with the HMS only and for one energy
with the protons
in coincidence in the DC. 
They are on average 2\% higher than the cross-sections extracted
from fits to the $ep$ data, which is consistent with the precision
of these data and of the present work.   
More details
on the experiment and on the data analysis will be found elsewhere \cite{DA}.

Our results are plotted in Figs.\ref{fig1} and \ref{fig2}. 
They smoothly approach the data from ALS \cite{PLA90}, 
but are in clear disagreement
with the lowest $Q^2$ point from SLAC \cite{ARN75}. 
Above 1 (GeV/c)$^2$,
our results are in good agreement with the SLAC data, significantly higher 
than the CEA \cite{ELI69} and Bonn \cite{CRA85} data. 
The CEA data were measured 
with  background contributions that might not have been subtracted 
reliably. 
>From Bonn, only the highest $Q^2$ point is a 
determination of $A$ independent of previous measurements. 
New results from a Jefferson Laboratory/Hall~A experiment also exist
\cite{HallA}.

Several classes of models can be used to calculate the deuteron form factors and
$A(Q^2)$. We will give only a few examples based on the most recent calculations.
For a review of earlier work, see e.g. \cite{PLA90,GAR94}.
All calculations are sensitive to the nucleon form factors, 
and in particular to
the poorly known neutron charge form factor ($G_E^n$)
\protect\cite{PLA90,SAR89,AMG98}. 
For instance, a 30\% change in $G_E^n$
results in about a 6\% change in $A$, in the $Q^2$ range of the data 
presented here.
$G_E^n$ should, however, be better determined from
several experiments in the near future. 
Different calculations for the deuteron form factors, including the
ones discussed below, do not use the same parameterization of the
nucleon form factors.
The non-relativistic impulse approximation (NRIA), 
using recent nucleon-nucleon potentials,
underestimates the structure function $A(Q^2)$. 
The addition of meson exchange currents (MEC)
and relativistic corrections (RC) improves the description, as illustrated by a
calculation using the Argonne $v_{18}$ potential \cite{WIR95}. 
There is however
some uncertainty about the pair term contribution \cite{AMG98} and about the
exact strength of the $\rho\pi\gamma$ contribution \cite{SAR89,MOS90}, so
that the good description given by the full non-relativistic calculation
of Ref. \cite{WIR95} still requires confirmation.
The new $A(Q^2)$ data should help fix these isoscalar MEC more accurately.
Note that most
nucleon-nucleon potentials, whether used in relativistic or non-relativistic
calculations, are adjusted to fit the nucleon-nucleon phase shifts 
up to only about 350 MeV, which might not be
high enough to match the $Q^2$ region of the present measurements. 
Isobar configurations (mostly $\Delta\Delta$) could also play a role in the
deuteron electromagnetic structure \cite{AMG98}.
For a better description of high $Q^2$ data, 
 several fully covariant approaches have been developed.
A solution of the Bethe-Salpeter equation
using the Gross approximation  (CIA) \cite{VAN95} and a calculation using 
light-front dynamics (LFD) \cite{CAR98} 
give similar results and are in reasonable agreement with our data.
They are, however, systematically too
high at lower $Q^2$, and differ at higher $Q^2$.
The LFD calculation \cite{CAR98} does not include the $\rho\pi\gamma$ MEC.
In quark-hadron hybrid models (QHM), 
 the addition of quark exchange terms leads to an overestimate
of $A(Q^2)$ \cite{BUC89}. Finally, 
parameterizing $A$ with a fall-off in $Q^{-2n}$, the exponent $n$
is found to increase from 2.7 to 4.5 in the region of this experiment,
still smaller than the value $n=10$ predicted asymptotically 
from quark counting rules \cite{PQCD}.

In conclusion, we have measured the $A(Q^2)$ deuteron structure function in
an intermediate momentum transfer  region. 
Our measurements resolve discrepancies between older experiments
and
yield significant constraints on the most recent and refined models
of the deuteron electromagnetic structure.
Their precision also requires a reexamination of small contributions such as
two-photon exchange.
Finally, they were performed at  the same kinematics
as polarization observables \cite{FUR98} and will
be used to extract the deuteron charge monopole and quadrupole form factors
 with good accuracy.

{\it Acknowledgements:} 
We acknowledge the outstanding work of the TJNAF accelerator division
and the Hall C engineering staff. 
We thank the Indiana University
Cyclotron Facility for its technical help with some components of the deuteron
channel. 
This work was supported by 
the French Centre National de la Recherche Scientifique and
the Commissariat \`a l'Energie Atomique,
the U.S. Department of Energy and
the National Science Foundation,
the Swiss National Science Foundation,
and the K.C. Wong Foundation.



\begin{table*}
\begin{tabular}{ccccccc} 
$Q^2$ (GeV/c)$^2$ & 0.657 &  0.786 & 1.017 & 1.178 & 1.510 & 1.790 \\ \hline 
$\theta_e$ (deg.) & 35.67 & 33.53 & 29.83 & 27.52 & 23.29 & 20.27	\\ \hline 
$d\sigma/d\Omega$ (nb/sr) & 77.1 & 43.3 & 19.4 & 11.6 & 5.20 & 2.75 \\ \hline	 
$B\times 10^{8}$ & 5090  & 2030  & 441  & 148  & 22.  & 3.6 	 \\ \hline 
$A\times 10^{6}$ & 323  & 194 & 88.3 & 51.9 & 20.9 & 9.77 	 \\ \hline 
$\Delta A_{\hbox{stat.}}/A$ (\%) & 0.6  & 0.8  & 0.9  & 0.8  & 0.9  & 1.1  \\\hline 
$\Delta A_{\hbox{syst.}}/A$ (\%) & 3.7  & 3.8  & 3.9   & 4.0  & 4.8  & 5.5  \\ 
\end{tabular}
\caption{Measured $ed$ cross-sections and extracted values of $A$, with statistical and
	 systematic errors. Values of $B$ are from a fit to the world data.}
\label{tab2}
\end{table*}
\begin{figure}
\caption{
	$A(Q^2)$: the data are from 
	previous experiments 
	and from this work . 
	The curves are from
	calculations \protect\cite{WIR95,VAN95,CAR98,BUC89}
	discussed in the text.
	}
\label{fig1}
\end{figure}
\begin{figure}[h]
\caption{
	$A(Q^2)$ deviation (in \%) from a fit
	to previous data,
	used as an arbitrary reference.
	For clarity, only previous data from Refs 
	\protect\cite{ARN75,CRA85,ELI69,PLA90} are shown. See Fig. 1
	for experiments and calculations legend. Our experimental
	error bars combine statistical and systematic errors in quadrature.
	}
\label{fig2}
\end{figure}
\newpage
\begin{figure}
\begin{center}
\mbox{\leavevmode \epsffile{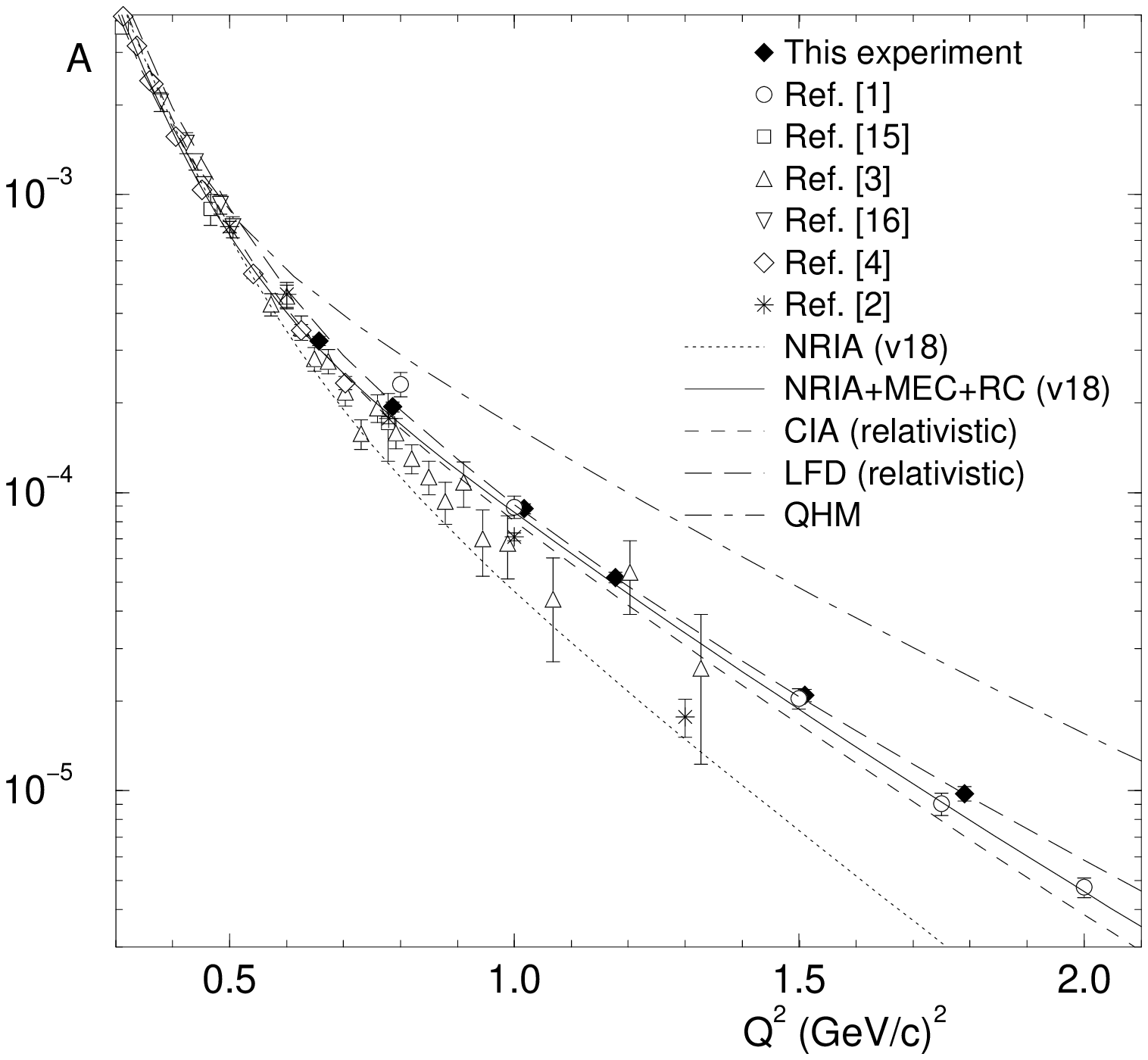}}
\end{center}
\end{figure}
\newpage
\hbox{\hfil}
\newpage
\begin{figure}
\begin{center}
\mbox{\leavevmode \epsffile{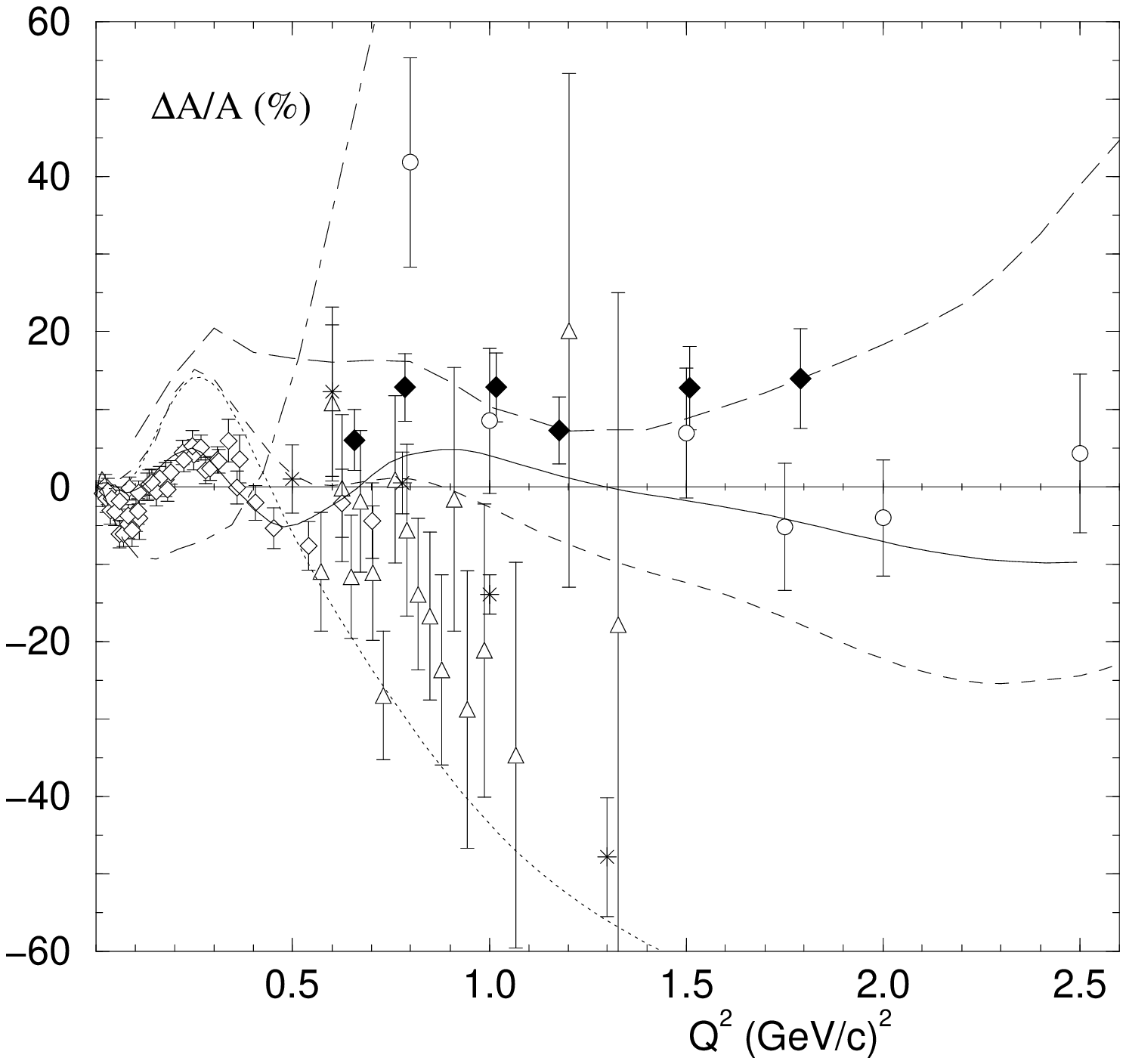}}
\end{center}
\end{figure}
\end{document}